\documentclass[conference,letterpaper]{IEEEtran}
\IEEEoverridecommandlockouts
\usepackage[utf8]{inputenc} 
\usepackage[T1]{fontenc}
\usepackage{url}
\usepackage{ifthen}
\usepackage{cite}
\usepackage[cmex10]{amsmath} 
\usepackage{amssymb}
\usepackage{graphicx}
\usepackage{epstopdf}
\usepackage{epsfig}
\usepackage{multirow}
\usepackage{amsthm}
\usepackage{comment}
\usepackage{caption}
\usepackage{enumitem}
\usepackage{amsmath}
\usepackage{color}
\usepackage{bbm}
\usepackage{algorithm}
\usepackage{algpseudocode}
\usepackage{csquotes}
\usepackage{subfiles}
\usepackage{caption, multirow, makecell}
\usepackage{array}
\usepackage{blkarray}
\usepackage{caption}
\usepackage{afterpage}
\usepackage{dblfloatfix}
\usepackage{hyperref}
\usepackage{graphicx}
\usepackage{balance}
\usepackage{subcaption}
\usepackage{diagbox}
\usepackage{setspace}
\usepackage{float}

\newtheorem{thm}{Theorem}
\newtheorem{lem}{Lemma}

\newtheorem{cor}{Corollary}
\newtheorem{example}{Example}

\newtheorem{defn}{Definition}

\newtheorem{rem}{Remark}

\usepackage{mathtools}
\newcommand\myeq{\stackrel{\mathclap{\normalfont\mbox{$\star$}}}{=}}

\def\BibTeX{{\rm B\kern-.05em{\sc i\kern-.025em b}\kern-.08em
		T\kern-.1667em\lower.7ex\hbox{E}\kern-.125emX}}
\begin{document}
	
	\title{On Hierarchical Coded Caching with Offline Users\\}
	
	\author{\IEEEauthorblockN{ Rashid Ummer N.T\IEEEauthorrefmark{1}, Charul Rajput\IEEEauthorrefmark{2}, B. Sundar Rajan\IEEEauthorrefmark{1}}
		\IEEEauthorblockA{\IEEEauthorrefmark{1}Department of Electrical Communication Engineering, Indian Institute of Science, Bengaluru, India
			\\\ \{rashidummer, bsrajan\}@iisc.ac.in}
		\IEEEauthorblockA{\IEEEauthorrefmark{2}Department of Mathematics and System Analysis, Aalto University, Finland
			\\\ charul.rajput@aalto.fi}}
		
	\maketitle
	
	\begin{abstract}
This paper studies a two-layer hierarchical network in which some users are offline during the content delivery phase. A two-layer hierarchical network consists of a single server connected to multiple cache-aided mirror sites, and each mirror site is connected to a distinct set of cache-aided users.  A scheme for such a hierarchical system with offline users has been proposed recently but considered a special case where all mirror caches have zero memory, which is a significant limitation. We propose an array known as a hierarchical hotplug placement delivery array (HHPDA), which describes the placement and delivery phases of a coded caching scheme for a general two-layer hierarchical network with offline users. Further, we construct a class of HHPDAs using combinatorial $t$-designs. 
	\end{abstract}
	\section{Introduction}
	Wireless network data traffic is steadily increasing, primarily due to the growing video demands \cite{Eri}. Cache-aided communication takes advantage of the temporal network traffic variability. Coded caching was first proposed by Maddah-Ali and Niesen in \cite{MaN} (referred to as MAN scheme), as a technique to reduce the peak traffic load in an error free broadcast network with multiple cache-aided users. The coded caching scheme operates in two phases: the \textit{placement phase} and the \textit{delivery phase}. In the placement phase during off-peak times, the server places contents in the user caches without knowing the future demands. In the delivery phase during peak times, the server broadcasts coded messages to satisfy the user's demands. In order to achieve the optimal load of the MAN scheme, each file needs to be split into packets, referred to as the subpacketization level, which grows exponentially with the number of users for a given memory ratio. Placement Delivery Array (PDA) was proposed as a tool to obtain low subpacketization level coded caching schemes for single-layer networks, by Yan \textit{et al.} in \cite{YCT}. Various PDA based coded caching schemes have been proposed in the literature \cite{Yan,ZCJ,MW,CWZW,Li,RS,WCLC,WCWC,WCWC2,CWWC}. 

	The MAN scheme and all other PDA based schemes require the file demands of all users to be known by the server in order to start the delivery. However, in a practical scenario, some users may fail to communicate their demands (we refer to such users as offline users) and thus the system fails to deliver the demands of active users. Coded caching with offline users (referred to as hotplug coded caching) was first studied by Ma \textit{et al.} in \cite{MT}. In a $(K, K', N)$ hotplug coded caching system, out of total $K$ users, only $K'$ users are active at the time of delivery. The number of active users needs to be known to the server during the placement phase. The hotplug coded caching schemes in \cite{MT} utilize the MAN scheme and thus require a subpacketization level that grows exponentially with the number of users. A combinatorial structure called Hotplug Placement Delivery Array (HpPDA) which describes a hotplug coded caching scheme was proposed in \cite{CS}. The authors in \cite{CS} showed that one existing hotplug scheme in \cite{MT} corresponds to a specific class of HpPDAs, and further improved the transmission load of that scheme. Construction of HpPDAs from combinatorial $t$-designs were discussed in \cite{CS_arx}. A new hotplug coded caching scheme from existing HpPDAs was given in \cite{MS}. Demand privacy in hotplug coded caching scheme was studied in \cite{MT2}.   
	
	Motivated by practical scenarios such as industrial Internet of Things networks, Karamchandani \textit{et al.} in \cite{KNMD} proposed a coded caching scheme for a two-layer hierarchical network, consisting of a single server with a library of $N$ files connected to $K_1$ mirrors each having a cache of size $M_1$ files and each mirror is connected to $K_2$ users each having a cache of size $M_2$ files.  Hierarchical coded caching schemes for the same network were also studied in \cite{KYWM}, \cite{ZZWXL}, \cite{WWCY} and \cite{RS_arx}. In \cite{KYWM}, Kong {\textit{et al.}} proposed an array called hierarchical placement delivery array (HPDA) that describes a hierarchical coded caching scheme. Low subpacketization level hierarchical schemes can be obtained by constructing appropriate HPDAs. HPDA constructions were discussed in \cite{KYWM} and \cite{RS_arx}. All the hierarchical coded caching schemes in \cite{KNMD,KYWM,ZZWXL,WWCY,RS_arx} require the file demands of all users to be known before the delivery phase. Hierarchical coded caching with offline users (referred to as hotplug hierarchical coded caching) has been studied recently in \cite{ACS}, but considered a special case where all mirror caches have zero memory, which is a significant limitation. In \cite{ACS}, the authors used existing HpPDAs to derive hotplug hierarchical coded caching schemes. It is of practical interest to study a general hotplug hierarchical coded caching system in which mirrors are also equipped with caches. All the hotplug schemes in \cite{MT,CS,CS_arx,MS,MT2} and \cite{ACS} uses coded placements using Maximum Distance Separable (MDS) codes \cite{LX2004}. An $[n, k]$ MDS code has the property that the information can be recovered from any set of $k$ code symbols out of $n$ symbols. 
	
	\textit{Contributions}: The contributions in this paper are as follows: \\
	\noindent $\bullet$ We introduce a combinatorial structure called Hierarchical Hotplug Placement Delivery Array (HHPDA), which describes the placement and delivery phases of a hotplug hierarchical coded caching scheme for a general two-layer hierarchical network. \\
	\noindent $\bullet$ Using combinatorial $t$-designs we construct a class of HHPDAs, that give hotplug hierarchical coded caching schemes for a general two-layer hierarchical network with non-zero mirror caches and user caches. 
	
	\textit{Organization}: The rest of the paper is organized as follows. In Section \ref{system}, the system model under consideration is described.  Section \ref{prelim_hhpda} reviews the concepts of PDA, HpPDA and HPDA. Useful definitions from combinatorial designs are also reviewed in this section. In Section \ref{hhpda_def}, we define HHPDA and explain the hotplug hierarchical coded caching scheme obtainable from a given HHPDA.  Section \ref{hhpda_tdes} covers the proposed construction of a class of HHPDAs from combinatorial $t$-designs. Section \ref{concl_hhpda} concludes the paper.
	
	\textit{Notations}: For $n,m \in \mathbb{Z}^{+}$, $[n]$ denotes the set $\{1,2,\ldots,n\}$ and $[n:m]$ denotes the set $\{n,n+1,n+2,\ldots, m\}$. For any integers $i$ and $n$ such that $0 \le i \le n$, $\binom{n}{i}$ denotes the binomial coefficient, calculated as $\frac{n!}{i!(n-i)!}$. For any set $\mathcal{A}$, $|\mathcal{A}|$ denotes the cardinality of $\mathcal{A}$. For a set $\mathcal{A}$ and a positive integer $i \leq |\mathcal{A}|$,  $\binom{\mathcal{A}}{i}$ denotes all the $i$-sized subsets of $\mathcal{A}$. For sets $\mathcal{A} \text{ and }\mathcal{B}$, $\mathcal{A}\textbackslash \mathcal{B}$ denotes the elements in $\mathcal{A}$ but not in $\mathcal{B}$.  \\
	\section{System model}\label{system}
\begin{figure}[!htbp]
		\centering
		\captionsetup{justification=centering}
		\includegraphics[width=0.48\textwidth]{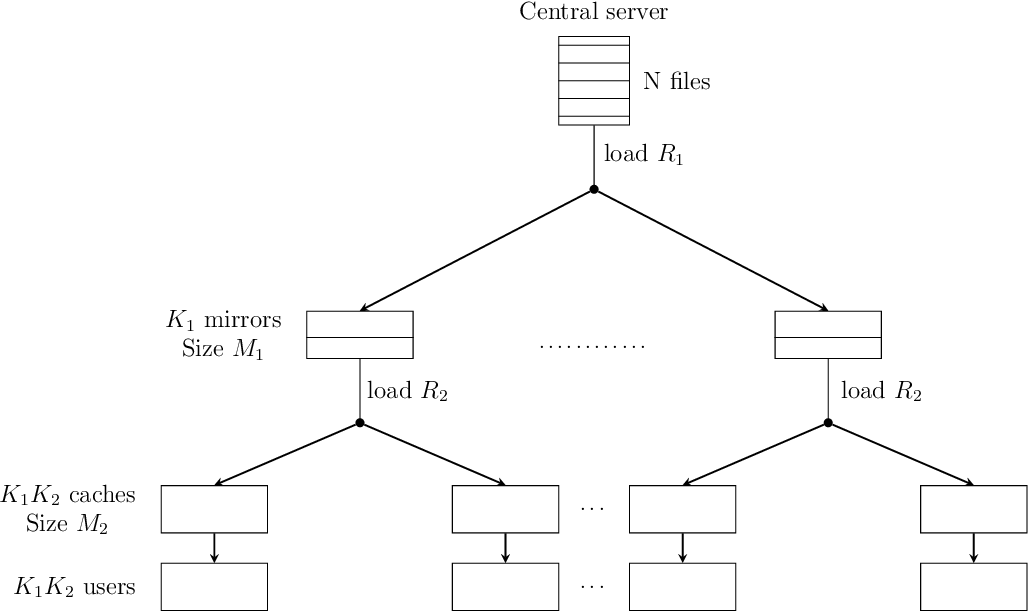}
		\caption{The two-layer $(K_1,K_2;M_1,M_2;N)$ hierarchical caching system.}
		\label{fig:setting2}
	\end{figure}

	A two-layer $(K_1, K_2; M_1, M_2; N)$ hierarchical caching system, depicted in Fig.\ref{fig:setting2}, comprises a central server with a library of $N$ files each of $B$ bits,  denoted by $\mathcal{W} = \{W_1, W_2, \dots, W_N\}$, connected via an error-free broadcast link to $K_1$ mirror sites, each equipped with a cache of size $M_1$ files. Each mirror site is further connected via an error-free broadcast link to $K_2$ users, each having a cache of size $M_2$ files. The transmission load from the server to the mirrors and from the mirrors to its attached users are denoted by $R_1$ and $R_2$, respectively. The set of users attached to the $k_1^{\text{th}}$ mirror site is denoted by $\mathcal{U}_{k_1}$, and the $k_2^{\text{th}}$ user in this set is denoted by $\mathcal{U}_{(k_1,k_2)}$. The cache contents of mirror site $k_1$ and user $\mathcal{U}_{(k_1,k_2)}$ are denoted by $\mathcal{Z}_{k_1}$ and $\mathcal{Z}_{(k_1,k_2)}$, respectively.  
	
	A two-layer $(K_1, K_2; M_1, M_2; N)$ hierarchical caching system in which $K'\le K_1K_2$ users are active is referred to as a $(K_1, K_2, K'; M_1, M_2; N)$ hotplug hierarchical caching system. The set of $K'$ active users is denoted by $\mathcal{U}' \subseteq \{(k_1,k_2) : k_1 \in [K_1], k_2 \in [K_2]\}, |\mathcal{U}'|=K'$. We arrange the users in $\mathcal{U}'$ in lexicographic order of $(k_1,k_2)$ and define a bijection $\phi(.)$ from $\mathcal{U}'$ to $[K']$. The set of active users attached to ${k_1}^{th}$ mirror is denoted as $\mathcal{U}'_{k_1}\subseteq \mathcal{U}'$. The set of requested files is represented by a demand vector $\vec{d}=(d_{1},d_{2},\ldots,d_{K'})$, where $d_{j}$ denotes the file demanded by $j^{th}$ user in $\mathcal{U}'$. The number of active users, not the identity, is assumed to be known to the server.
	\section{Preliminaries}\label{prelim_hhpda}
	In this section, we first review the concepts of PDA, HpPDA and HPDA. Some useful definitions and properties from combinatorial designs that are used in this paper are also reviewed.
\subsection{Placement Delivery Array (PDA)}
\begin{defn} {\normalfont{\cite{YCT}}}
	For positive integers $K, F, Z$ and $S$, an $F \times K$ array $\mathbf{P}=(p_{j,k})_{j \in [F], k \in [K]}$, composed of a specific symbol $\star$ and $S$ non-negative integers $1,2,\ldots,S$, is called a $(K, F, Z, S)$ PDA if it satisfies the following conditions: \\
	\textit{C1}. The symbol $\star$ appears $Z$ times in each column.\\
	\textit{C2}. Each integer occurs at least once in the array.\\
	\textit{C3}. For any two distinct entries $p_{j_1,k_1}$ and $p_{j_2,k_2}$, we have $p_{j_1,k_1}=p_{j_2,k_2}=s$ is an integer only if (\textit{a}) $j_1 \neq j_2$, $k_1 \neq k_2$, i.e., they lie in distinct rows and distinct columns, and (\textit{b}) $p_{j_1,k_2}=p_{j_2,k_1}=\star$.
\end{defn} 
\begin{lem}{\normalfont{\cite{YCT}}}\label{lem:pda_scheme}
	For a given $(K, F, Z, S)$ PDA $\mathbf{P}=(p_{j,k})_{F \times K}$, a $(K,M,N)$ coded caching scheme can be obtained with subpacketization $F$ and $\frac{M}{N}=\frac{Z}{F}$. Any possible demands from all users are met with a load of $R=\frac{S}{F}$.
\end{lem}
\subsection{Hotplug Placement Delivery Array (HpPDA)}
\begin{defn}{\normalfont{\cite{CS}}}
	Let $K, K', F, F`, Z, Z'$ and $S$ be integers such that $K \geq K', F \geq F'$ and $Z<F'$. Consider two arrays given as follows
	\begin{itemize}
		\item $\mathbf{P}=(p_{f,k})_{f \in [F], k \in [K]}$ which is an array containing `$\star$' and null. Each column contains $Z$ number of stars.
		\item $\mathbf{B}=(b_{f,k})_{f \in [F'], k \in [K']}$ which is a $(K', F', Z', S)$ PDA.	
	\end{itemize}
	For each $\tau \subseteq [K], |\tau| = K'$, there exists a subset $\zeta \subseteq [F], |\zeta|=F'$ such that 
	\begin{equation*}
		[\mathbf{P}]_{\zeta \times \tau} \myeq B,
	\end{equation*}	
	where $[\mathbf{P}]_{\zeta \times \tau}$ denotes the subarray of $P$ whose rows corresponds to the set $\zeta$ and columns corresponds to the set $\tau$, and $\myeq$ denotes that the positions of all stars are same in both the arrays. This is referred to as a $(K,K',F, F',Z,Z',S)$- HpPDA $(\mathbf{P}, \mathbf{B})$.
\end{defn}
\begin{lem}\label{thm2.1}\cite{CS}
Given a $(K,K',F, F',Z,Z',S)$-HpPDA $(P, B)$, there exists a $(K, K', N)$ hotplug coded caching scheme with subpacketization level $F'$, memory ratio $\frac{M}{N}=\frac{Z}{F'},$ where $M$ denotes the number of files each user can store in it's cache and transmission load $R=\frac{S}{F'}$, in which $[F, F']$ MDS code is used for encoding the subfiles of each file.
\end{lem}
\subsection{Hierarchical Placement Delivery Array (HPDA)}
\begin{defn}\cite{KYWM}
For any given positive integers $K_1, K_2, F, Z_1, Z_2$ with $Z_1 < F$, $Z_2 < F $ and any integer sets $S_m$ and $S_{k_1}, k_1 \in [K_1]$, an  $ F \times (K_1+K_1K_2)$ array $\mathbf{Q}= \left( \mathbf{Q^{(0)}},\mathbf{Q^{(1)}},\ldots,\mathbf{Q^{(K_1)}} \right)$, where $\mathbf{Q^{(0)}}=(q_{j,k_1}^{(0)})_{j \in [F], k_1 \in [K_1]}$, is an $F \times K_1$ array consisting of $\star$ and \textit{null}  and $\mathbf{Q^{(k_1)}}=(q_{j,k_2}^{(k_1)})_{j \in [F], k_2 \in [K_2]}$ is an $F \times K_2$  array over  $\{ \star \cup S_{k_1} \}, k_1 \in [K_1]$, is a $(K_1, K_2;F;Z_1,Z_2;S_m,S_1, \ldots, S_{K_1})$ hierarchical placement delivery array  if it satisfies the following conditions: \\
	\textit{B1}. Each column of $\mathbf{Q^{(0)}}$ has $Z_1$ stars.\\
	\textit{B2}. $\mathbf{Q^{(k_1)}}$ is a $(K_2,F,Z_2,|S_{k_1}|)$ PDA for each $ k_1 \in [K_1]$ .\\
	\textit{B3}. Each integer $s \in S_m$ occurs in exactly one subarray $\mathbf{Q^{(k_1)}}$ where $ k_1 \in [K_1]$ and for each $(q_{j,k_2}^{(k_1)})=s \in S_m,j \in [F],k_1 \in [K_1], k_2 \in [K_2]$, we have $(q_{j,k_1}^{(0)})=\star$. \\
	\textit{B4}. For any two entries $q_{j,k_2}^{(k_1)}$ and $q_{j',k'_2}^{(k'_1)}$, where $k_1 \neq k'_1 \in [K_1]$, $j,j' \in [F]$ and $k_2,k'_2 \in [K_2]$, if $q_{j,k_2}^{(k_1)}=q_{j',k'_2}^{(k'_1)}$ is an integer then
	\begin{enumerate}[label=-]		
		\item $q_{j',k_2}^{(k_1)}$ is an integer only if $q_{j',k_1}^{(0)}=\star$,
		\item $q_{j,k'_2}^{(k'_1)}$ is an integer only if $q_{j,k'_1}^{(0)}=\star$.\\
	\end{enumerate}
\end{defn}  
	\begin{lem}{\normalfont{\cite{KYWM}}}\label{lem:hpda}
	Given a $(K_1,K_2;F;Z_1,Z_2;S_m,S_1,.., S_{K_1})$  HPDA $\mathbf{Q}= \left( \mathbf{Q^{(0)}},\mathbf{Q^{(1)}}, \ldots, \mathbf{Q^{(K_1)}} \right)$, a $(K_1,K_2;M_1,M_2;N)$ coded caching scheme is obtained with subpacketization level $F$, memory ratios  $\frac{M_1}{N}=\frac{Z_1}{F}$, $\frac{M_2}{N}=\frac{Z_2}{F}$ and transmission loads   $R_1=\frac{\left|\underset{k_1=1}{\bigcup^{K_1} S_{k_1}}\right|-\left|S_m \right|}{F}$  and  $R_2=\max_{k_1 \in [K_1]}\left\{ \frac{|S_{k_1}|}{F}\right\}$.
\end{lem}

\subsection{Combinatorial Designs }\label{designs}
\begin{defn}[Design $(\mathcal{X}, \mathcal{A})$ \cite{Stin}]
	A design is a pair  $(\mathcal{X}, \mathcal{A})$ such that the following properties are satisfied: 
	\textit{1}. $\mathcal{X}$ is a set of elements called points, and
	\textit{2}. $\mathcal{A}$ is a collection (i.e., multiset) of nonempty subsets of $\mathcal{X}$ called blocks.
\end{defn}
\begin{defn}[$t-(v, k, \lambda)$ design \cite{Stin}]\label{def:tdes}
	Let $v,k,\lambda$ and $t$ be positive integers such that $v>k\geq t$. A $t-(v, k, \lambda)$ design is a design $(\mathcal{X}, \mathcal{A})$ such that the following properties are satisfied: \\
	\textit{1}. $|\mathcal{X}|=v$, 
	\textit{2}. each block contains exactly $k$ points, and 
	\textit{3}. every set of $t$ distinct points is contained in exactly $\lambda$ blocks.
\end{defn}
 The existence and construction of $t$-designs  have been discussed in \cite{Stin}, \cite{Col}, \cite{RW} and \cite{TT}. For convenience, we write blocks of $t$-designs in the form $abcd$ rather than $\{a,b,c,d\}$.
\begin{example}\label{ex:example1}
	$\mathcal{X}=\{ 0,1,2,3,4,5,6,7,8,9 \}$, $ \mathcal{A}=\{ 0123,0145,0246,0378,0579,0689,1278,1369,1479,1568, \\ 2359,2489,2567,3458,3467 \}$ is a $2$-$(10,4,2)$ design. 
\end{example}
The following Lemmas will be used in the construction of HHPDAs in Section \ref{hhpda_def}.		
\begin{lem}{\normalfont\cite{Stin}}\label{lem:tdes}
	Suppose that $(\mathcal{X}, \mathcal{A})$ is a $t-(v, k,\lambda)$ design. Suppose that $\mathcal{Y} \subseteq \mathcal{X}$, where $|\mathcal{Y}|=s\leq t$. Then there are exactly $\lambda_{s}=\frac{\lambda \binom{v-s}{t-s}}{\binom{k-s}{t-s}}$ blocks in $\mathcal{A}$ that contains all the points in $\mathcal{Y}$. Thus the number of blocks in $\mathcal{A}$, $b=\lambda_{0}=\frac{\lambda \binom{v}{t}}{\binom{k}{t}}$.
\end{lem}
The following is an immediate corollary of Lemma \ref{lem:tdes}.
\begin{cor}\label{cor:tdes}
	Suppose that $(\mathcal{X}, \mathcal{A})$ is a $t-(v, k,\lambda)$ design, and $1 \le s \le t$. Then $(\mathcal{X}, \mathcal{A})$ is an $s-(v, k,\lambda_s)$ design, where $\lambda_{s}=\frac{\lambda \binom{v-s}{t-s}}{\binom{k-s}{t-s}}$. 
\end{cor}	
	\begin{lem}{\normalfont\cite{Stin}}\label{lem:tdes2}
	Let $(\mathcal{X}, \mathcal{A})$ be a $t-(v, k, \lambda)$ design. Suppose that $\mathcal{Y} \subseteq \mathcal{Z} \subseteq \mathcal{X}$, where $ |\mathcal{Y}|=i, |\mathcal{Z}|=j$, and $i \le j$. Then there are exactly $\lambda_{i}^{t}=\frac{\lambda \binom{v-j}{k-i}}{\binom{v-j}{k-j}}$ blocks in $\mathcal{A}$ that contains all the points in $\mathcal{Y}$ and none of the points in $\mathcal{Z}$. 
\end{lem}

\section{Hierarchical Hotplug Placement Delivery Array}\label{hhpda_def}
In this section, we first introduce a combinatorial structure called Hierarchical Hotplug Placement Delivery Array (HHPDA). Then, we show that corresponding to any HHPDA, there exist a hotplug hierarchical coded caching scheme. 
\begin{defn}[Hierarchical Hotplug Placement Delivery Array]\label{def:hhpda}
Consider any positive integers $K_1, K_2, K', F, F', Z_1, Z_2, Z'$ with $Z_1+Z_2 < F' \le F$, $K' \le K_1K_2$,  and any integer sets $\mathcal{S}$ and $\mathcal{S}_{k_1}, k_1 \in [K_1]$, and $\mathcal{S} \cap \mathcal{S}_{k_1} = \emptyset$. Consider two arrays given as follows. \\
\noindent $\bullet$ An  $ F \times (K_1+K_1K_2)$ array $\mathbf{Q}= \left( \mathbf{Q^{(0)}},\mathbf{Q^{(1)}}, \ldots, \mathbf{Q^{(K_1)}} \right)$, where $\mathbf{Q^{(0)}}=(q_{f,k_1}^{(0)})_{f \in [F], k_1 \in [K_1]}$, is an $F \times K_1$ array consisting of $\star$ and \textit{null},  and $\mathbf{Q^{(k_1)}}=(q_{f,k_2}^{(k_1)})_{f \in [F], k_2 \in [K_2]}$ is an $F \times K_2$  array over  $(\{\star\} \cup \{\textit{ null }\} \cup \mathcal{S}_{k_1} ), k_1 \in [K_1]$, satisfies the following conditions: \\
	\textit{A1}. Each column of $\mathbf{Q^{(0)}}$ has $Z_1$ stars. \\
	\textit{A2}. Each column of $\mathbf{Q^{(k_1)}}$ has $Z_2$ stars, for each $k_1 \in [K_1]$. \\
	\textit{A3}. Each integer $s \in \mathcal{S}_{k_1}$ occurs in exactly the subarray $\mathbf{Q^{(k_1)}}$ where $ k_1 \in [K_1]$, and if $q_{f,k_1}^{(0)}=\star$, then $q_{f,k_2}^{(k_1)} \in \{\star\} \cup \mathcal{S}_{k_1} $, $\forall$ $f \in [F],k_1 \in [K_1], k_2 \in [K_2]$. \\
	\textit{A4}.  For any two distinct entries $q_{f,k_i}^{(k_1)}$ and $q_{f',k_j}^{(k_1)}$, we have $q_{f,k_i}^{(k_1)}=q_{f',k_j}^{(k_1)}=s \in \mathcal{S}_{k_1}$ is an integer only if (\textit{a}) $f \neq f'$, $k_i \neq k_j$, 
and (\textit{b}) $q_{f,k_j}^{(k_1)}=q_{f',k_i}^{(k_1)}=\star$. \\
	\noindent $\bullet$ An $ F' \times K'$ array $\mathbf{B}=(b_{f',k})_{f' \in [F'], k \in [K']}$ which is a $(K', F', Z', |\mathcal{S}|)$ PDA.	\\
For each $\tau \subseteq \{(k_1, k_2) | k_1 \in [K_1], k_2\in [K_2]\}, |\tau| = K'$, there exists a subset $\zeta \subseteq [F], |\zeta|=F'$ such that 
\begin{equation}\label{eq:hhpda}
	\mathbf{Q'}_{\zeta,\tau} \myeq \mathbf{B},
\end{equation}
where array $\mathbf{Q'}_{\zeta,\tau} = (q'_{f', (k_1, k_2)})_{f' \in \zeta, (k_2,k_2) \in \tau}$ is an $F' \times K' $ array defined as 
	\begin{equation}\label{arrayQ}
		q'_{f', (k_1, k_2)}=\begin{cases}
			\star & \text{if} \text{ either } q_{f',k_1}^{(0)}=\star \text{ or } q_{f',k_2}^{(k_1)}=\star \\
			\text{null} & \text{ otherwise, }
		\end{cases}
	\end{equation}
and $\myeq$ denotes that the positions of all stars are same in both the arrays. This is referred to as a $(K_1,K_2,K';F, F';Z_1,Z_2,\\Z';\mathcal{S},\mathcal{S}_1, \ldots,\mathcal{S}_{K_1})$ HHPDA $(\mathbf{Q}, \mathbf{B})$.
\end{defn}
Next, we provide an example of HHPDA.
\begin{example}\label{hhpda_ex1}
	It is easy to see that the following arrays $(\mathbf{Q}, \mathbf{B})$  form a $(4,2,3;14, 9;3,4,5;\mathcal{S}=[5],\mathcal{S}_1=\{6:11\},\mathcal{S}_2=\{12:17\},\mathcal{S}_3=\{18:23\},\mathcal{S}_4=\{24:29\})$ HHPDA. 
	
{\footnotesize	
\[ \mathbf{Q}=
\begin{blockarray}{cccccccccccc}
\begin{block}{[cccc|cc|cc|cc|cc]}
 \star &\star & &        & 6 & 7 & 12 & 13 &   &   &   &   \\
 \star & & \star &         & 8 & 9 &   &   &  18 & 19 &   &   \\
     \star & & & \star        & 10 & 11 &   &   &   &   & 24 & 25 \\
    & & &                                                                   & \star &   & \star &   & \star &   & \star &   \\
      & & &                                                              & \star &   & \star &   &   & \star &   & \star \\
      & & &                                                             & \star &   &   & \star & \star &   &   & \star \\
     & & &                                                        & \star &   &   & \star &   & \star & \star &   \\
   &\star & &\star         &   &   & 14 & 15 &   &   & 26 & 27 \\
       & & &                                                      &   & \star &   & \star &   & \star &   & \star \\
      & & &                                                   &   & \star & \star &   & \star &   &   & \star \\
      & & &                                                        &   & \star & \star &   &   & \star & \star &   \\
     & & &                                                             &   & \star &   & \star & \star &   & \star &   \\
     &\star & \star &         &   &   & 16 & 17 & 20 & 21 &   &   \\
 & & \star & \star        &   &   &   &   & 22 & 23 & 28 & 29 \\
\end{block}
\end{blockarray}
\]
and
\[
\mathbf{B} = \begin{bmatrix}
         \star & \star & 1 \\
         \star & 1 & \star \\
         1 & \star & \star \\
         \star & \star & 2 \\
         \star & 2 & \star \\
         2 & \star & \star \\
         \star & 3 & 4 \\
         3 & \star & 5 \\
         4 & 5 & \star  \\
\end{bmatrix}.
\]
}

\end{example}

In the following theorem, we show that corresponding to any 
HHPDA $(\mathbf{Q}, \mathbf{B})$, there exist a coded caching scheme for a  hotplug hierarchical caching system.

\begin{thm}\label{thm:hhpda}
Given a $(K_1,K_2,K';F, F';Z_1,Z_2,Z';\mathcal{S},\mathcal{S}_1, \\ \ldots, \mathcal{S}_{K_1})$ HHPDA $(\mathbf{Q}, \mathbf{B})$, 
a $(K_1, K_2, K'; M_1, M_2; N)$ hotplug hierarchical coded caching scheme can be obtained using Algorithm \ref{alg:hhpda} with subpacketization level $F'$, memory ratios  $\frac{M_1}{N}=\frac{Z_1}{F'}$, $\frac{M_2}{N}=\frac{Z_2}{F'}$ and transmission loads   $R_1=\frac{\left|\mathcal{S}\right|}{F'}$  and  $R_2=\frac{1}{F'} \displaystyle \max_{k_1 \in [K_1]}\left\{ |\underset { \tiny \begin{array}{c} j \in [K'], \\ \phi^{-1}(j)\in \mathcal{U}'_{k_1} \end{array}}{\bigcup } \hspace{-2mm} \mathcal{S}^{(j)} |+| \underset { \tiny \begin{array}{c} k_2 \in [K_2], \\ (k_1, k_2)\in \mathcal{U}'_{k_1} \end{array}}{\bigcup } {\hspace{-2mm}\mathcal{S}_{k_1}^{(k_2,\zeta)}}|\right\}$, 
where the set of $K'$ active users is denoted by $\mathcal{U}'\subseteq \{(k_1,k_2) : k_1 \in [K_1], k_2 \in [K_2]\}, |\mathcal{U}'|=K'$ arranged in the lexicographic order of $(k_1,k_2)$, and $\phi(.)$ is a bijection from $\mathcal{U}'$ to $[K']$. Let $\mathcal{U}'_{k_1} \subseteq \mathcal{U}'$ is the set of active users attached to mirror $k_1$. The set $\mathcal{S}^{(j)} \subseteq \mathcal{S}$ is the set of integers appearing in the $j^{th}$ column of $\textbf{B}$. The set ${\mathcal{S}_{k_1}^{(k_2,\zeta)}} \subseteq \mathcal{S}_{k_1}$ is the set of integers appearing in the $k_2^{th}$ column and the rows indexed by $f \in \zeta$ of array $\mathbf{Q^{(k_1)}}$.
\end{thm}
The proof of Theorem \ref{thm:hhpda} is given in \textit{Appendix \ref{appendix:hhpda_proof}}.

\begin{algorithm}[]
	\renewcommand{\thealgorithm}{1}
	\caption{Hotplug hierarchical coded caching scheme based on a HHPDA $(\mathbf{Q}=\left( \mathbf{Q^{(0)}},\mathbf{Q^{(1)}}, \ldots,\mathbf{Q^{(K_1)}} \right), \mathbf{B})$}
	\label{alg:hhpda}
	\begin{algorithmic}[1]
		\Procedure{Placement}{$\mathbf{Q},\mathcal{W}$}       
		\State Split each file $W_n \in \mathcal{W}$ into $F'$ packets, i.e., $W_n =\{W_{n,f'} | f' \in [F']\}$.  Encode $F'$ packets of each file $W_n, n \in [N]$, into $F$ coded packets using $[F, F']$ MDS code with generator matrix $G$ of order $F' \times F$, i.e.,
		$$\begin{bmatrix}
			C_{n,1} \\ C_{n,2} \\ \vdots \\ C_{n, F} 
		\end{bmatrix} =  G^{T}\begin{bmatrix}
			W_{n,1} \\ W_{n,2} \\ \vdots \\ W_{n, F'} 
		\end{bmatrix} .$$ 
		\For{\texttt{$k_1 \in [K_1]$}}
		\State  $\mathcal{Z}_{k_1}$ $\leftarrow$ $\{C_{n,f}: q_{f,k_1}^{(0)}=\star, n \in [N],f \in [F]\}.$
		\EndFor
		\For{\texttt{$(k_1,k_2), k_1 \in [K_1], k_2 \in [K_2] $}}
		\State $\mathcal{Z}_{(k_1,k_2)}$ $\leftarrow$ $\{C_{n,f}: q_{f,k_2}^{(k_1)}=\star, n \in [N],f \in [F]\}.$
		\EndFor
		\EndProcedure	
\Procedure{Delivery Server}{$\mathbf{Q},\mathbf{B},\mathcal{W},\mathcal{U}',\vec{d} \hspace{0.1cm}$}
		\State 	For a set of active user $\mathcal{U}'$, find a subset $\zeta \subseteq [F], |\zeta|=F'$ such that $\mathbf{Q'}_{\zeta, \mathcal{U}'} \myeq \mathbf{B} $, where $\mathbf{Q'}_{\zeta, \mathcal{U}'}$ is defined in \eqref{arrayQ}.
		\State Make a new array $\overline{\mathbf{Q}}=(\overline{q}_{f,k})_{f \in \zeta, k \in [K']}$ by filling $s \in \mathcal{S}$ integers in null spaces of the array $\mathbf{Q'}_{\zeta, \mathcal{U}'}$ in such a way that $\overline{\mathbf{Q}}=\mathbf{B}$.
		\For{\texttt{$s \in \mathcal{S} $}}
		\State Server sends the following coded message to the mirror sites:
		\State $X_s=\underset{\overline{q}_{f,k}=s, f\in \zeta,k\in[K']}{\bigoplus}C_{d_{k},f}$
		\EndFor    
		\EndProcedure	
		\Procedure{Delivery Mirror $k_1$}{$\mathbf{Q},\mathbf{B},\mathcal{W},\mathcal{U}',\vec{d}, X_s $} 
		\State Let $\mathcal{S}^{(j)} \subseteq \mathcal{S}$ is the set of integers appearing in the $j^{th}$ column of $\textbf{B}$.
		\For{$ s \in \mathcal{S}^{(j)}$ for which $\phi^{-1}(j) \in \mathcal{U}'_{k_1} $}
		\State After receiving $X_s$, mirror site $k_1$ sends the following coded message to users in $\mathcal{U}_{k_1}$:
		\State \hspace{-0.4cm} $X_{k_1,s}=X_s{\bigoplus}\left(\tiny \underset{\begin{array}{c}  q^{(0)}_{f,k_1}=\star, \overline{q}_{f,k'}=s, f\in \zeta, \\ k' \in [K'] \backslash \phi(\mathcal{U}'_{k_1}) \end{array}}{\bigoplus}C_{d_{k'},f}\right)$
		\EndFor
		\State  Let $\mathcal{S}_{k_1}^{(k_2,\zeta)} \subseteq \mathcal{S}_{k_1}$ be the set of integers appearing in $k_2^{th}$ column and the rows indexed by $f \in \zeta$ of $\mathbf{Q^{(k_1)}}$.
		\For{$s' \in \mathcal{S}_{k_1}^{(k_2,\zeta)},$ for which $(k_1,k_2) \in \mathcal{U}'_{k_1}, $}
		\State Mirror site $k_1$ sends the following coded signal:
		\State $X_{k_1,s'}=\underset{q_{f,k_2}^{(k_1)}=s', f\in \zeta, \phi((k_1,k_2))=j  }{\bigoplus}C_{d_{j},f}$
		\EndFor   
		\EndProcedure	
	\end{algorithmic}
\end{algorithm}
\begin{rem}
	In Definition \ref{def:hhpda} of HHPDA, if $Z_1=0$, then discarding $\mathbf{Q^{(0)}}$ from $\mathbf{Q}$, results in an array $(\mathbf{Q}, \mathbf{B})$ that satisfies the definition of HpPDA. The resulting scheme corresponds to the hotplug hierarchical scheme described in \cite{ACS}.
\end{rem}

The following example illustrates Theorem \ref{thm:hhpda} and Algorithm \ref{alg:hhpda}.
	\begin{example}\label{hhpda_ex2}
	Consider the $(4,2,3;14, 9;3,4,5;\mathcal{S}=[5],\mathcal{S}_1=\{6:11\},\mathcal{S}_2=\{12:17\},\mathcal{S}_3=\{18:23\},\mathcal{S}_4=\{24:29\})$ HHPDA $(\mathbf{Q}, \mathbf{B})$ in Example \ref{hhpda_ex1}. Based on this HHPDA, one can obtain a $(4,2,3;\frac{N}{3},\frac{4N}{9};N)$ hotplug hierarchical coded caching scheme using Algorithm \ref{alg:hhpda} as follows. \\
	\noindent $\bullet$ \textit{Placement phase:} From line $2$ of Algorithm \ref{alg:hhpda}, each file $W_n, \forall n \in [N]$ is split into $9$ packets, i.e., $W_n =\{W_{n,1},W_{n,2}, \ldots, W_{n,9}\}$. These packets are then encoded using a $[9,14]$ MDS code. The $14$ coded packets of each file is denoted by $\{C_{n,1},C_{n,2}, \ldots,C_{n,14}\}$. By line $3-5$ of Algorithm \ref{alg:hhpda}, the packets cached in each mirror site is given by, $\mathcal{Z}_1 = \{C_{n,1},C_{n,2},C_{n,3}\}, \mathcal{Z}_2 = \{C_{n,1},C_{n,8},C_{n,13}\}, \mathcal{Z}_3 = \{C_{n,2},C_{n,13},C_{n,14}\}, \mathcal{Z}_4 = \{C_{n,3},C_{n,8},C_{n,14}\}, \forall n \in [N]$. Therefore, $\frac{M_1}{N}=\frac{Z_1}{F'}=\frac{3}{9}=\frac{1}{3}$. By lines $6-8$ of Algorithm \ref{alg:hhpda}, four coded packets are cached by each user. For example, the packets cached by user $\mathcal{U}_{(2,1)}$ is given by, $\mathcal{Z}_{(2,1)} = \{C_{n,4},C_{n,5},C_{n,10},C_{n,11}\}, \forall n \in [N]$. Therefore, $\frac{M_2}{N}=\frac{Z_2}{F'}=\frac{4}{9}$.\\
	\noindent $\bullet$ \textbf{Delivery phase}: Since $K'=3$, any three users are active during delivery phase. Let  $\mathcal{U}'=\{{(1,1)},{(2,2)},{(3,1)}\}$ be the set of $3$ active users arranged in the lexicographic order of $(k_1,k_2)$. Consider a bijection $\phi(.)$ from $\mathcal{U}'$ to $[3]$. Then the subset $\zeta \subseteq [14], |\zeta|=9$ such that $\mathbf{Q'}_{\zeta, \mathcal{U}'} \myeq \mathbf{B} $, specified in line $11$ of Algorithm \ref{alg:hhpda}, is given by $\zeta =\{1,2,12,7,4,13,3,8,14\}$. Therefore, by line $12$ of Algorithm \ref{alg:hhpda}, one can construct the array $\mathbf{Q'}_{\zeta, \mathcal{U}'}$ such that $\overline{\mathbf{Q}}=\mathbf{B}$. That is,
	\begin{equation*}
		\overline{\mathbf{Q}}=\begin{blockarray}{cccc}
			& 1 & 2 & 3  \\
			\begin{block}{c[ccc ]}
				1 &\star & \star &  1 \\
				2 & \star & 1 & \star\\
				12 &1 & \star & \star\\
				7 &\star & \star & 2  \\
				4 & \star & 2 & \star\\
				13 & 2 & \star & \star\\
				3 &\star & 3 &  4 \\
				8 & 3 & \star & 5 \\
				14 &4 & 5 & \star\\
			\end{block}
		\end{blockarray}.
	\end{equation*}
	
	Let demand vector $\vec{d}=(d_{1},d_{2},d_3)=(W_1,W_2,W_3)$. Then by lines $13-16$, the server sends the following coded messages. 
	$ X_1 = C_{1,12} \oplus C_{2,2} \oplus C_{3,1} , X_2 = C_{1,13} \oplus C_{2,4} \oplus C_{3,7}, X_3 = C_{1,8} \oplus C_{2,3}, X_4 = C_{1,14} \oplus C_{3,3} \text{ and } X_5 = C_{2,14} \oplus C_{3,8}$.
	Therefore the load of the first layer is $R_1=\frac{5}{9}$. 
	
	After receiving $X_1$, $X_2$, $X_3$, $X_4$ and $X_5$,  by lines $19-23$ of Algorithm \ref{alg:hhpda}, the mirror site $k_1=1$ sends the coded messages $X_{1,s}$ as follows.
	$X_{1,1}=X_1{\bigoplus} C_{2,2} \oplus C_{3,1} = C_{1,12}$, $X_{1,2}=X_2$, $X_{1,3}=X_3$, $X_{1,4}=X_4$. 
	Similarly, mirror site $2$ sends $X_{2,1}=X_1{\bigoplus} C_{3,1} =C_{1,12} \oplus C_{2,2}$, $X_{2,2}=X_2 $, $X_{2,3}=X_3{\bigoplus}C_{1,8} = C_{2,3}$, $X_{2,5}=X_5{\bigoplus}C_{3,8} =C_{2,14}$. 
	Similarly, mirror site $3$ sends $X_{3,1}=X_1{\bigoplus} C_{2,2} =C_{1,12} \oplus C_{3,1}$, $X_{3,2}=X_2 {\bigoplus}C_{1,13}= C_{2,4} \oplus C_{3,7}$, $X_{3,4}=X_4{\bigoplus}C_{1,14} = C_{3,3}$, $X_{3,5}=X_5{\bigoplus}C_{2,14} =C_{3,8}$. Therefore, $R'_2=\frac{4}{9}$. 
	
	By lines $24-28$ of Algorithm \ref{alg:hhpda}, the mirror sites $1$ sends the messages $X_{1,6}=C_{1,1}, X_{1,8}=C_{1,2}$ and $X_{1,9}=C_{1,3}$.  Similarly, mirror sites $2$ sends the messages $X_{2,13}=C_{2,1}, X_{2,15}=C_{2,8}$ and $X_{2,17}=C_{2,13}$. Similarly, mirror sites $3$ sends the messages $X_{3,18}=C_{3,2}, X_{3,20}=C_{3,13}$ and $X_{3,22}=C_{3,14}$. Therefore, $R''_2=\frac{3}{9}$. Therefore the load of the second layer is $R_2=R'_2+R''_2=\frac{7}{9}$.
	Each user can thus recover the demanded file. For example, consider user $(1,1)$. User  $(1,1)$ has in its cache the packets $C_{1,4}, C_{1,5}, C_{1,6}$ and $C_{1,7}$. From the transmissions $X_{1,1}, X_{1,2}, X_{1,3}$ and $X_{1,4}$,  user $(1,1)$ gets the packets  $C_{1,12}, C_{1,13}, C_{1,8}$ and $C_{1,14}$, respectively. From the transmissions $X_{1,6}, X_{1,8}$ and $X_{1,9}$,  user $(1,1)$ gets the packets  $C_{1,1}, C_{1,2}$ and $C_{1,3}$, respectively. Thus it has $9$ packets out of $14$ coded packets of the demanded file $W_{1}$. Since we have used a $[7,14]$ MDS code, the user can decode the demanded file $W_{1}$ correctly.

\end{example}
\begin{figure*}[ht]  
	\centering	
	\begin{equation*}	
		{\scriptsize
			\mathbf{Q}=
			\setlength{\arraycolsep}{2pt}
			\begin{blockarray}{ccccccccccccc} 
				&12 & 34 & 56 & 78 & 1 & 2 & 3 & 4 & 5 & 6 & 7 & 8 \\
				\begin{block}{c[cccc|cccccccc]}
					1234 &  \star &\star & &        & 234 & 134 &  124 & 123 &   &   &   &   \\
					1256 &   \star & & \star &         & 256 & 156 &   &   & 126 & 125 &   &   \\
					1278 &    \star & & & \star       & 178 & 278 &   &   &   &   & 128 & 127 \\
					1357 &    & & &                                                                   & \star &   & \star &   & \star &   & \star &   \\
					1368 &      & & &                                                              & \star &   & \star &   &   & \star &   & \star \\
					1458 &      & & &                                                             & \star &   &   & \star & \star &   &   & \star \\
					1467 &     & & &                                                        & \star &   &   & \star &   & \star & \star &   \\
					3478 &   &\star & &\star         &   &   & 478 & 378 &   &   & 348 & 347 \\
					2468 &       & & &                                                      &   & \star &   & \star &   & \star &   & \star \\
					2358 &       & & &                                                   &   & \star & \star &   & \star &   &   & \star \\
					2367 &      & & &                                                        &   & \star & \star &   &   & \star & \star &   \\
					2457 &      & & &                                                             &   & \star &   & \star & \star &   & \star &   \\
					3456 &     &\star & \star &         &   &   & 456 & 356 & 346 & 345 &   &   \\
					5678 &   & & \star & \star        &   &   &   &   & 678 & 578 & 568 & 567 \\
				\end{block}
			\end{blockarray},
			\hfill
			\mathbf{B}=\begin{blockarray}{cccc}
				& 1 & 2 & 3  \\
				\begin{block}{c[ccc ]}
					(12, 1) &\star & \star &  (123, 1) \\
					(13, 1) & \star & (123, 1) & \star\\
					(23, 1) &(123, 1) & \star & \star\\
					(12, 2) &\star & \star & (123, 2)  \\
					(13, 2) & \star & (123, 2) & \star\\
					(23, 2) & (123, 2) & \star & \star\\
					(1, 1) &\star & (12, 1) &  (13, 1) \\
					(2, 1) & (12, 1) & \star & (23, 1) \\
					(3, 1) &(13, 1) & (23, 1) & \star\\
				\end{block}
			\end{blockarray}.
		}
	\end{equation*}	
	\caption{Arrays $(\mathbf{Q}, \mathbf{B})$ in Example \ref{hhpda_tdes_ex}.}
	\label{fig:hhpda}	
\end{figure*}	
\section{A class of HHPDAs using $t$-Designs}\label{hhpda_tdes}
In this section, we first construct a class of HHPDAs using $t$-designs, and then the parameters of that class are given in Theorem \ref{thmt}. 

\noindent\textbf{Construction:}
Let $(X,\mathcal{A})$ be a $t$-$(v,k,\lambda)$ design with non-repeated blocks, where $X=\{1,2,\ldots, v\}$, $\mathcal{A}=\{A_1, A_2,$ $\ldots, A_b\}$ and $|A_i|=k$ for all $i \in \{1,2, \ldots, b\}$. Let $K_2$ be an integer such that $K_2<t$, and $v=K_1K_2$. Consider an array $\mathbf{Q}= \left( \mathbf{Q^{(0)}},\mathbf{Q^{(1)}}, \ldots, \mathbf{Q^{(K_1)}} \right)$ whose rows are indexed by the blocks in $\mathcal{A}$. Define a set $\mathcal{D}=\{D_1, D_2, \ldots, D_{K_1}\}$, where $D_{k_1}=\{(k_1-1)K_2+1, \ldots, k_1K_2\}$ for all $k_1 \in [K_1]$. The array $\mathbf{Q^{(0)}}=(q_{A,D}^{(0)})_{A \in \mathcal{A}, D \in \mathcal{D}}$, is a $b \times K_1$ array defined as
\begin{equation} \label{arrayQ0}
q_{A,D}^{(0)}=\begin{cases}
\star & \text{if} \ \ D \subseteq A, \\
null &  \text{otherwise}
\end{cases}.
\end{equation}
The array $(\mathbf{Q^{(1)}}, \ldots, \mathbf{Q^{(K_1)}})=(q_{A, i})_{A \in \mathcal{A}, i \in X}$ is an $b \times v$  array defined as 
\begin{equation} \label{arrayQk1}
q_{A,i}=\begin{cases}
\star & \text{if} \ \ i \in A \ \& \ q_{A,D_j}^{(0)} \ne \star, \ \text{for} \ i \in D_j \\
A\backslash\{i\} & \text{if} \ \ i \in A \ \& \ q_{A,D_j}^{(0)} = \star, \ \text{for} \ i \in D_j \\
null &  \text{otherwise}
\end{cases}.
\end{equation}

 The array $\mathbf{B}$ is defined in same manner as in the $t$-scheme given in \cite{CS_arx}. For $1 \leq s \leq t-1$, let $0 \leq a_s \leq \lambda_s^{t}$, and consider a set 
$\mathcal{\mathcal{R}}=\bigcup_{s=1}^{t-1} \left\{ (Y, i) \mid Y \in {[t] \choose s}, i \in [a_s] \right\}.$
Clearly, $|\mathcal{R}|=\sum_{s=1}^{t-1} a_s {t \choose s}$. The integers $a_s\in \{0,1, \ldots, \lambda_s^t\}, s \in [t-1]$ are chosen in such a way that $|\mathcal{R}| > \lambda_1$. 
Consider an array $B$ whose rows are indexed by the elements in $\mathcal{R}$ and columns are indexed by the points in $[t]$. The array $\mathbf{B}=(B((Y, i), j))$ is a $|\mathcal{R}| \times t$ array, defined as
\begin{equation} \label{arrayB}
B((Y, i), j)=\begin{cases}
* & \text{if} \ \ j \in Y, \\
\left(Y\cup\{j\},i\right) & \text{if} \ \ j \notin Y
\end{cases}.
\end{equation}
\begin{example}\label{hhpda_tdes_ex} 
 For a $3$-$(8, 4, 1)$ design, with $b=14$, and choosing $a_2 = 2$ and $a_1 = 1$, we get the arrays $(\mathbf{Q}, \mathbf{B})$ shown in Fig.\ref{fig:hhpda}. If we replace all the entries represented as different sets by different integers, we get the HHPDA $(\mathbf{Q}, \mathbf{B})$ given in Example \ref{hhpda_ex1}.
\end{example}

The following theorem gives the parameters of a class of HHPDAs constructed above using a $t$-design.
\begin{thm}\label{thmt}
    Given a $t$-$(v, k, \lambda)$ design with $b$ number of blocks, 
there exists a $(K_1,K_2,K';F, F';Z_1,Z_2,Z';\mathcal{S},\mathcal{S}_1, \ldots,\mathcal{S}_{K_1})$ HHPDA  $(\mathbf{Q}, \mathbf{B})$ with $K_2 \leq t$,  and  for some $s \in [t-1],$ $a_s \in \{0, 1, \ldots , \lambda_s^t \}$,
\begin{align*}
& K_1K_2=v, K'=t, F=b, F'=\sum_{s=1}^{t-1} a_s {t\choose s}, \\
& Z_1=\lambda_{K_2}, Z_2=\lambda_1-\lambda_{K_2}, Z'=\sum_{s=1}^{t-1} a_s {t-1\choose s-1}, \\
&|\mathcal{S}|=\sum_{s=1}^{t-1} a_s {t\choose s+1}, |\mathcal{S}_{k_1}|=\lambda_{k_2} K_2 \ \text{for all} \ k_1 \in [K_1].
\end{align*}
\end{thm}
The proof of Theorem \ref{thmt} is given in Appendix \ref{appendix_hhpda_tdes}.
The following corollary directly holds from Theorems \ref{thm:hhpda} and \ref{thmt}.
\begin{cor}
    Given a $t$-$(v, k, \lambda)$ design with $b$ number of blocks, there exists a scheme for a $(K_1, K_2, K'; M_1, M_2; N)$ hotplug hierarchical coded caching scheme with $K_1K_2=v$, $K'=t$ and $K_2 \leq t$, 
If $\mathcal{U}'=\cup_{k_1 \in [K_1]} \mathcal{U}'_{k_1}$ is the set of active users, where $\mathcal{U}'_{k_1}$ denotes the set of active users attached to mirror $k_1$, $k_1 \in [K_1]$, then the parameters of the scheme are given as follows: $\frac{M_1}{N}  = \frac{\lambda_{K_2}}{F'}$, $\frac{M_2}{N}  = \frac{\lambda-\lambda_{K_2}}{F'}$, $R_1  = \frac{|\mathcal{S}|}{F'}$, $R_2 =  \frac{1}{F'}  \underset{k_1 \in [K_1]} \max \left\{ | \cup_{j \in \mathcal{U}'_{k_1}} \mathcal{S}^{(j)} | +| \mathcal{U}'_{k_1}|\alpha \right\}$, where $\alpha \leq \lambda_{K_2}$, 
$F'$ and $|S|$ is same as defined in Theorem \ref{thmt}, and $\mathcal{S}^{(j)}$ is the set of integers appearing in the $j$-th column of $\mathbf{B}$.

\end{cor}
\section{Conclusion}\label{concl_hhpda}
In this paper, we considered a two-layer hierarchical caching network in which some users are offline during content delivery. We introduced a combinatorial structure called HHPDA which gives a hotplug hierarchical coded caching scheme. By using combinatorial $t$-designs, we then constructed a class of HHPDAs. Obtaining hotplug hierarchical schemes, for arbitrary numbers of mirrors and users, and for any given mirror and user cache sizes, remains an interesting open problem.

\section*{Acknowledgement}
This work was supported partly by the Science and Engineering Research Board (SERB) of Department of Science and Technology (DST), Government of India, through J.C Bose National Fellowship to B. Sundar Rajan.

	\begin{appendices}
	\section{Proof of Theorem \ref{thm:hhpda}}\label{appendix:hhpda_proof}
	We show that from a given $(K_1,K_2,K';F, F';Z_1,Z_2,Z';\\\mathcal{S},\mathcal{S}_1, \ldots,\mathcal{S}_{K_1})$ HHPDA $(\mathbf{Q},\mathbf{B})$,  Algorithm \ref{alg:hhpda} gives a $(K_1, K_2, K'; M_1, M_2; N)$ hotplug hierarchical coded caching scheme. The scheme operates in two phases. 
	
	$\bullet$ \textit{Placement phase:} From lines $1$ to $3$ of Algorithm \ref{alg:hhpda} it is clear that, in the placement phase the server first splits each file into $F'$ packets, i.e., $W_n =\{W_{n,f'} | f' \in [F']\}, \forall n \in [N]$. Therefore, the subpacketization level is $F'$. Then the server encodes $F'$ packets of each file into $F$ coded packets using $[F, F']$ MDS code with generator matrix $G$ of order $F' \times F$, i.e.,   
	$$\begin{bmatrix}
		C_{n,1} \\ C_{n,2} \\ \vdots \\ C_{n, F} 
	\end{bmatrix} =  G^{T}\begin{bmatrix}
		W_{n,1} \\ W_{n,2} \\ \vdots \\ W_{n, F'} 
	\end{bmatrix} .$$ 
	The server fills the cache of $k_1^{th}$ mirror by coded packets according to line $5$ of Algorithm \ref{alg:hhpda}, i.e.,  $\mathcal{Z}_{k_1}$ $\leftarrow$ $\{C_{n,f}: q_{f,k_1}^{(0)}=\star, n \in [N],f \in [F]\}$. Since, each column of $\mathbf{Q^{(0)}}$ has $Z_1$ stars and the size of each coded packet is $\frac{B}{F'}$, we have $\frac{M_1}{N}=\frac{Z_1}{F'}$. The server fills the cache of user $\mathcal{U}_{(k_1,k_2)}$ by coded packets according to line $8$ of Algorithm \ref{alg:hhpda}, i.e.,  $\mathcal{Z}_{(k_1,k_2)}$ $\leftarrow$ $\{C_{n,f}: q_{f,k_2}^{(k_1)}=\star, n \in [N],f \in [F]\}$. Since, $\mathbf{Q^{(k_1)}}$ has $Z_2$ stars, and the size of each coded packet is $\frac{B}{F'}$, we have $\frac{M_2}{N}=\frac{Z_2}{F'}$.
	
	$\bullet$ \textit{Delivery phase:}  Let the demand vector $\vec{d}=(d_{1},d_{2},\ldots,d_{K'})$. In the delivery phase, the server first broadcast coded messages to the mirrors, as given in lines $10$ to $17$ of Algorithm \ref{alg:hhpda}. By Definition \ref{def:hhpda} of HHPDA $(\mathbf{Q},\mathbf{B})$, for any given set of active users $\mathcal{U}'$, one can find a subset $\zeta \subseteq [F], |\zeta|=F'$ such that $\mathbf{Q'}_{\zeta,\mathcal{U}'} \myeq B $ and the entries of $\mathbf{Q'}$ satisfy (\ref{arrayQ}). Then, a new array $\overline{\mathbf{Q}}=(\overline{q}_{f,k})_{f \in \zeta, k \in [K']}$ is obtained by filling $s \in \mathcal{S}$ integers in null spaces of the subarray $[\mathbf{Q'}]_{\zeta \times \mathcal{U}'}$ in such a way that $\overline{\mathbf{Q}}=\mathbf{B}$.  The array $\overline{\mathbf{Q}}$ is used for transmissions from the server to the mirrors. Since $\mathbf{B}$ is a $(K', F', Z', |\mathcal{S}|)$ PDA and $\overline{\mathbf{Q}}=\mathbf{B}$, the number of integers in $\overline{\mathbf{Q}}$ is equal to $|\mathcal{S}|$. For each $s \in \mathcal{S}$, the server sends the coded message, $X_s=\underset{\overline{q}_{f,k}=s, f\in \zeta,k\in[K']}{\bigoplus}C_{d_{k},f}$. Since each transmission is of a packet size, $R_1=\frac{|\mathcal{S}|}{F'}$ files. 
	
	The set of $K'$ active users is denoted by $\mathcal{U}'\subseteq \{(k_1,k_2) : k_1 \in [K_1], k_2 \in [K_2]\}, |\mathcal{U}'|=K'$ arranged in the lexicographic order of $(k_1,k_2)$, and $\phi(.)$ is a bijection from $\mathcal{U}'$ to $[K']$. Let $\mathcal{U}'_{k_1} \subseteq \mathcal{U}'$ is the set of active users attached to mirror $k_1$. The mirror transmissions are represented by two set of integers. Let $\mathcal{S}^{(j)} \subseteq \mathcal{S}$ is the set of integers appearing in the $j^{th}$ column of $\textbf{B}$. By lines $19-23$ of Algorithm \ref{alg:hhpda}, after receiving $X_s$, for each $ s \in \mathcal{S}^{(j)}$ for which $\phi^{-1}(j) \in \mathcal{U}'_{k_1} $, mirror site $k_1$ sends the coded message $X_{k_1,s}$ given by,\\ $X_{k_1,s}=X_s{\bigoplus}\left(\tiny \underset{\begin{array}{c}  q^{(0)}_{f,k_1}=\star, \overline{q}_{f,k'}=s, \\ f\in \zeta,  k' \in [K'] \backslash \phi(\mathcal{U}'_{k_1}) \end{array}}{\bigoplus}C_{d_{k'},f}\right)$. That is, from the received message $X_s$, the $k_1^{th}$ mirror cancels the coded packets $C_{d_{k'},f}$, which are packets corresponding to the demands of users attached to mirrors other than $k_1$ and also present in mirror $k_1$'s cache. From (\ref{arrayQ}) and since $\overline{\mathbf{Q}}$ is a PDA, it is clear that if a packet corresponding to an integer $ s \in \mathcal{S}^{(j)}$ other than the packet of the demanded file is not present in the associated mirror cache, then it will be in the user's cache. Therefore, for each active user connected to mirror $k_1$, $X_{k_1,s}$ contains one packet required by that user and the remaining packets in $X_{k_1,s}$ are present in the user's cache. Let $R'_2$ denotes the mirror load corresponding to these transmissions. Then, $R'_2=\frac{1}{F'} \displaystyle \max_{k_1 \in [K_1]}\left\{ |\underset { \tiny \begin{array}{c} j \in [K'], \phi^{-1}(j)\in \mathcal{U}'_{k_1} \end{array}}{\bigcup } \hspace{-2mm} \mathcal{S}^{(j)} |\right\}$. From the above transmissions, user $\phi^{-1}(j)\in \mathcal{U}'_{k_1}$ gets $|\mathcal{S}_j|=F'-Z'$ coded packets of the demanded file. 
	
	 Now consider the rows in which $\star$'s appear in a given column $j$ of $\overline{\mathbf{Q}}$. Out of the $Z'$ $\star$'s in each column of $\overline{\mathbf{Q}}$, consider the row indices $f' \in \zeta$ of the $Z_1$ rows in which $\star$'s appear such that $(q_{f',k_1}^{(0)})=\star$ in  $\mathbf{Q^{(0)}}$. Then by properties \textit{A3} and  \textit{A4} of Definition \ref{def:hhpda}, $(q_{f,k_2}^{(k_1)}) \in \{\star, \mathcal{S}_{k_1} \}$.  Let ${\mathcal{S}_{k_1}^{(k_2,\zeta)}} \subseteq \mathcal{S}_{k_1}$ denote the set of integers appearing in the $k_2^{th}$ column and the rows indexed by $f \in \zeta$ of array $\mathbf{Q^{(k_1)}}$.  Then, by lines $24-29$, for each  $s' \in \mathcal{S}_{k_1}^{(k_2,\zeta)}$ for which $(k_1,k_2) \in \mathcal{U}'_{k_1}$,  mirror site $k_1$ sends the coded signal $X_{k_1,s'}=\underset{q_{f,k_2}^{(k_1)}=s', f\in \zeta, \phi((k_1,k_2))=j  }{\bigoplus}C_{d_{j},f}$. By this set of transmissions together from its cache contents, user $\phi^{-1}(j)\in \mathcal{U}'_{k_1}$ is ensured to get $Z'$ packets of the demanded file. Let $R^{''}_2$ denotes the mirror load corresponding to these transmissions. Then, $R^{''}_2=\frac{1}{F'}  \displaystyle \max_{k_1 \in [K_1]}| \underset { \tiny \begin{array}{c} k_2 \in [K_2], (k_1, k_2)\in \mathcal{U}'_{k_1} \end{array}}{\bigcup } {\hspace{-2mm}\mathcal{S}_{k_1}^{(k_2,\zeta)}}|$. Therefore, the transmissions corresponding to $ s \in \mathcal{S}^{(j)}$ and $s' \in \mathcal{S}_{k_1}^{(k_2,\zeta)}$ together ensures that at least $F'$ packets of the demanded file is obtained by each user in $\mathcal{U}'$.  Since the packets are coded using $[F, F']$ MDS code, each user can decode the demanded file correctly from the coded subfiles it have. The overall transmission load $R_2= R^{'}_2 + R^{''}_2 =\frac{1}{F'} \displaystyle \max_{k_1 \in [K_1]}\left\{ |\underset { \tiny \begin{array}{c} j \in [K'], \\ \phi^{-1}(j)\in \mathcal{U}'_{k_1} \end{array}}{\bigcup } \hspace{-2mm} \mathcal{S}^{(j)} |+| \underset { \tiny \begin{array}{c} k_2 \in [K_2], \\ (k_1, k_2)\in \mathcal{U}'_{k_1} \end{array}}{\bigcup } {\hspace{-2mm}\mathcal{S}_{k_1}^{(k_2,\zeta)}}|\right\}$. \hfill $\blacksquare$

\section{Proof of Theorem \ref{thmt}}\label{appendix_hhpda_tdes}
In this section we prove that the arrays $\mathbf{Q}$ and $\mathbf{B}$ constructed in \eqref{arrayQ0}, \eqref{arrayQk1} and \eqref{arrayB} forms a HHPDA with the parameters given in Theorem \ref{thmt}.

	\textit{A1}. Since $|D_{k_1}|=K_2$ for all $k_1 \in K_1$ and any set of $K_2$ points in a $t$-$(v,k,\lambda)$ design belongs to exactly $\lambda_{K_2}$ blocks, we have $Z_1=\lambda_{K_2}$.

\textit{A2}. The number of blocks containing a point $i \in X$ is $\lambda_{1}$. From \eqref{arrayQk1}, we have
$q_{A,i}=\star \ \text{if} \  i \in A \ \& \ q_{A,D_j}^{(0)} \ne \star, \ \text{for} \ i \in D_j $, therefore $Z_2=\lambda_1 -Z_1=\lambda_1 - \lambda_{K_2}$.

\textit{A3}. To prove this point, we show that each integer in array $\mathbf{Q}$ appears only once. Assume that there are two integer entries $A\backslash\{i\}$ and $A'\backslash\{i'\}$ in array $\mathcal{Q}$  for some $A, A' \in \mathcal{A}$ and $i,i' \in X$ such that 
\begin{equation}\label{eqA}
A\backslash\{i\}=A'\backslash\{i'\}.
\end{equation}
 From \eqref{arrayQk1}, we know $i\in A$ and $i' \in A'$. If $i\neq i'$, then equality in \eqref{eqA} is not possible. Therefore $i=i'$, then from \eqref{eqA}, we have $A=A'$, which means both entries are same.  

\textit{A4.} Since we proved that each integer in array $\mathbf{Q}$ appears only once, condition \textit{A4} becomes trivially true.

Since the array $\mathbf{B}$ is defined in same manner as in $t$-scheme given in \cite{CS_arx}, we refer to Lemma 3 in \cite{CS_arx} for the proof that $\mathbf{B}$ is a $(K', F', Z', S)$ PDA.

Now we need to prove condition \eqref{eq:hhpda} for which we construct array $\mathbf{P}=(p_{A, i})_{A\in\mathcal{A}, i \in X}$ using $\mathbf{Q}$ as
\begin{equation} \label{arrayP}
p_{A,i}=\begin{cases}
null & \text{if} \ \ q_{A,i} =null, \\
\star &  \text{otherwise}
\end{cases},
\end{equation}
which is equivalent to 
\begin{equation*} 
p_{A,i}=\begin{cases}
\star & \text{if} \ \ i \in A, \\
null &  \text{otherwise}
\end{cases}.
\end{equation*}
Clearly, $p_{A,i}= \star$ if either $q^{(0)}_{A,D_j} = \star$ for $i \in D_j$ or $q_{A, i} =\star$. By the construction of HpPDA given in  \cite{CS_arx}, $(\mathbf{P},\mathbf{B})$ is an HpPDA, which implies that for any set $\tau \subseteq [K_1K_2], |\tau|=K'$, there exists a set $\zeta \subseteq \mathcal{A}, |\zeta|=F'$ such that $[\mathbf{P}]_{\zeta \times \tau} \myeq \mathbf{B}$. This subarray $[\mathbf{P}]_{\zeta \times \tau}$  is equivalent to $\mathbf{Q}'_{\zeta \times \tau}=(q'_{A, k'})_{A\in \mathcal{A}, k' \in[K']}$, i.e.,
\begin{equation*} 
q'_{A, k'}=[p]_{A,k'}=\begin{cases}
\star & \text{if either}  \ q^{(0)}_{A,D_j} = \star, \ \text{for} \  k' \in D_j \\
    &  \text{or} \ q_{A,k'} =\star, \\
null &  \text{otherwise}
\end{cases}.
\end{equation*}
Hence $(\mathbf{Q},\mathbf{B})$ is a HHPDA. \hfill $\blacksquare$

	\end{appendices}
		
\end{document}